\documentclass[11pt]{amsart}
\usepackage{amsmath,amssymb,amscd}
\usepackage{amsmath,amssymb}
\usepackage{mathrsfs}
\usepackage{color}
\usepackage{amsxtra, amsmath}
\usepackage{amssymb, amscd}
\usepackage{graphicx}
\theoremstyle{plain}

\theoremstyle{definition}

\theoremstyle{remark}

\title[Bispectrality and the ad-conditions]{Bispectrality and the ad-conditions}

\author{F. A.  Gr\"unbaum}

\address{Department of Mathematics, University of California, Berkeley
CA 94705}
\email{grunbaum@math.berkeley.edu}
\subjclass[2010]{33C45, 22E45, 33C47}

\keywords{The bispectral problem, The ad-conditions, Darboux's process, Exceptional orthogonal polynomials, Matrix valued orthogonal polynomials}

\begin{document}

\begin{abstract} At the beginning of the study of the bispectral problem, see \cite{DG}, the ad-conditions played a crucial role in finding non-classical instances. The connection with the ad-conditions has reappeared in several different incarnations of the bispectral problem.
Here we show that, properly adapted versions of these conditions, see \cite{Reach123}, can play an important role in areas including, for instance, the study of exceptional orthogonal polynomials. This is also true in the non-commutative case. Even at this more advanced stage of the field one may hope that finding explicit solutions of these ad-conditions will provide an additional route to new examples.

%The non-commutative situation was started by a towering figure: M. G. Krein. Most of his work is closely tied to deep applications of mathematics, and the bispectral problem neither in the scalar or matrix valued case has yet found applications. Looking for methods than can furnish new examples could help in this search.  

\end{abstract}

\maketitle

\section{A brief introduction}

We recall that the bispectral problem posed and solved in \cite{DG} consists of finding potentials $V$ such that some family of eigenfunction $\psi(x,k)$ of 
the Schr\"odinger operator $\mathcal{L}$ given by

$$ \mathcal{L} \psi(x,k)=(-\dfrac{d^2}{dx^2} + V(x)) \psi(x,k)= k^2 \psi(x,k)$$

\noindent
should be eigenfunctions of a differential operator of finite order $m$ in the spectral variable $k$ with an eigenvalue $\Theta(x)$, i.e. there should exist a differential operator $\mathcal{B}(k,d/dk)$ such that

$$ \mathcal{B}(k,d/dk) \psi(x,k) = \Theta(x) \psi(x,k)$$
 
\bigskip

 The starting point of \cite{DG} was the observation that the existence of a bispectral situation for the Schr\"odinger operator $\mathcal{L}$ given by

\begin{equation}\label{Schr}
-\dfrac{d^2}{dx^2} + V(x).
\end{equation}

\noindent
is {\bf equivalent} to having a solution to the "ad-conditions", i.e. the differential operator 
 ${ad}\ \mathcal{L}^{m+1}(\Theta)$ obtained by taking repeated commutators of $\mathcal{L}$ (of order $2$) with $\Theta(x)$ (of order zero), should vanish identically, giving

\begin{equation}\label{ad} 
 \textrm{ad}\ \mathcal{L}^{m+1}(\Theta)=0.
\end{equation}

%`where we use the standard notation
%``\begin{equation}\label{ads}
%\textrm{ad}\ X(Y)=XY-YX 
%\end{equation}for the usual commutator and \begin{equation}\label{adgeneral}\left(\textrm{ad}\ X\right)^n(Y)=\textrm{ad}\ X \left(\left(\textrm{ad}\ X\right)^{n-1}(Y)\right),\quad  n\geq 2.\end{equation}

\bigskip
	It is not immediately obvious why this Lie algebraic nilpotency condition should be the answer to this unrelated question. In particular it implies that $V(x)$ needs to be rational, something that may be rather hard to see directly.

\bigskip

	The proof of this equivalence consists of two parts: the necessity of (2) is given in a few lines on page $181$, the converse is a much more elaborate argument given in the rest of page $181$ and all of page $182$ of \cite{DG}. 

\bigskip

Well known classical examples of this bispectral situation are given by the Bessel and Airy operators, with $m=2$.
	The first new examples of a bispectral situation in \cite{DG} came about by painfully solving equation (2) above when $m=4$ and $\Theta$ an arbitrary polynomial of degree $4$.

\bigskip

The observation that one family of solutions was given by the rational solutions $V$ of the Korteweg-deVries equation, previously found in
\cite{AiMcMos}, gave us a clue that many other pieces of magic could be floating around: the importance of the Darboux process in producing new solutions out of old ones as established 
in \cite{AdMos}, the role of characters of certain representations of $GL(N)$ as discovered by the school of M. Sato, see \cite{Sato}, the relation with trivial or very simple monodromy first observed in \cite{DG}, see also 
\cite{GonchVeselov},
	\cite{Oblom}. For a very recent look at this issue see \cite{FeldVa}. 

\bigskip

In \cite{DG} a second family of solutions was also given explicitly, and it was later observed in \cite{ZM} that these involve the so called "master symmetries" of the KdV equation. The KdV flows are obtained by repeated application of the Nijenhuis tensor acting on translations. Its master symmetries are obtained by replacing translations by dilations, as nicely explained in \cite{ZM}.
	These two families exhaust all non-classical solutions, as shown in \cite{DG}.

\bigskip

A consequence of these facts is that the solutions of the bispectral problem in the continuous-continuous case of \cite{DG}, organize themselves in nice smooth manifolds where very interesting vector fields span their tangent spaces. Nothing like this appears to hold in the discrete-continuous case to be considered in this paper. We will return to this issue later on.

\bigskip

	Some form of the "ad-conditions" appeared in several variants of the continuous-continuous version of the problem considered in \cite{DG}. The Berkeley PhD thesis of Michael Reach, (1987) (with results reported in \cite{Reach123,R,R1}) where he considered a continuous-discrete version of the original problem, as well as the papers \cite{GH0,GRUNHAIN} with Luc Haine, where we considered a q-version of the problem are the first examples after \cite{DG} showing that a bispectral situation is usually linked to some form of the ad-conditions. See also \cite{GH1}, where we insist on a three term recursion but allow for a differential operator of arbitrary order, and obtains conditions of the same kind as in \cite{DG}. In contrast with \cite{DG} in these other cases there is no claim of a strict equivalence between a bispectral situation and some form of the "ad-conditions", but there is a one way implication.

\bigskip

	The ad-conditions have been considered in the non-commutative case, see \cite{GT,GI,G5,CG08}, at times in conjunction with the Darboux process. In particular \cite{GT} contains a proof independent of \cite{DG} of the equivalence mentioned earlier for the non-commutative case.For more recent work on this line, see \cite{CGYZ4} and \cite{BoDuZu}. In this paper we find a way of connecting the ad-conditions with a different take on the bispectral problem obtained by exchanging the role for the physical and spectral variables.

\bigskip

	I cannot emphasize enough the importance of the ad-conditions for the results obtained in \cite{DG}; without them it is hard to imagine how we could have gone beyond the known cases with $m=2$. We were also very lucky in posing the problem  for $\mathcal{L}$ of order $2$. Although many of the considerations hold for $\mathcal{L}$ of higher order, solving equation (2) would have been hopeless when the order is higher than $2$. See \cite{W1} for some general results. One more piece of good fortune: considering a more general $\mathcal{L}$ would have led to the general Kadomtsev-Petviashvilii family of flows, beyond the more familiar KdV flows. For a reference to KdV, KP and related flows, see for instance \cite{NMPZ}

\bigskip

We use below the abbreviation 

$$ A_j=\textrm{ad}\ \mathcal{L}^{j}(\Theta)$$

\noindent
and it may be worth emphasizing that $A_j$ is a differential operator of order $j$ acting on smooth enough functions.

\bigskip

	In \cite{Reach123} one finds a very ingenious argument, vaguely inspired in \cite{DG}. It is given in two lemmas in section $6$ of \cite{Reach123}, showing that in the case when the eigenfunctions of the differential operator satisfy a $2 m+1$ recursion relation and the spectrum of the differential operator $\mathcal{L}$ is linear in $n$ (such as in the Hermite and Laguerre cases) there exist universal constants $a_j$ such that the following {\bf operator identity} holds

\begin{equation}\label{mike}
\sum_{i=0}^{2 m +1} a_{i} A_i= 0
\end{equation}

When the spectrum is quadratic in $n$, as in the Jacobi case, a more elaborate relation among several powers of
$ \textrm{ad}\ \mathcal{L}^j(\Theta)$ and
	$\mathcal{L}$ itself holds. This is an observation of M. Reach in \cite{Reach123}.

\bigskip

In the case of a $3$ term recursion relation satisfied by the eigenfunctions $\psi_n$
	the argument of M. Reach alluded to above, gives a first equation of the kind given in (3) above, namely

\begin{equation}\label{equ}
	A_3-A_1=0
\end{equation}

\noindent
see $(10)$ in \cite{Reach123}. This equation is fully solved in Reach's paper and gives two possible families of solutions going with the Hermite and the generalized Laguerre polynomials. These are the unique solutions up to translation and dilation.

\bigskip

	The next example of an ad-condition that follows from the method of M. Reach mentioned above, assuming that the eigenfunctions of $\mathcal{L}, \psi_n$, satisfy a $5$ term recursion relation is

\begin{equation}
A_5- 5 A_3 + 4 A_1=0
\end{equation}

\noindent
This is the most elaborate ad-condition given by M. Reach in \cite{Reach123,R1}, and our purpose is to show how his ideas can be pushed further.

\bigskip

	To be quite explicit: the two ad-conditions above, (4) and (5), assume that the spectrum of the differential operator $\mathcal{L}$ is given by $\lambda_n=n$ as in the Laguerre case. In the Hermite case when the spectrum is $\lambda_n=2 n$ they have to be modified to

\begin{equation}
A_3-4 A_1=0
\end{equation}

and 

\begin{equation}
A_5- 20 A_3 + 64 A_1=0
\end{equation}

\bigskip

We are now ready to state the new results of this paper.

\section{The contents of the paper}

	Recall that in \cite{DG} we obtained the first interesting examples by solving explicitly the ad-conditions (2) for $\mathcal{L}$ and $\Theta(x)$. As mentioned above these ad-conditions were crucial in making some headway at that early stage of the game.
As mentioned above, M. Reach found in his continuous-discrete version of the problem the general solution of his first ad-condition 
$$A_3-A_1=0$$
\noindent
and using this he showed that Laguerre and Hermite( up to some scaling) are the only cases going with $3$ term recursions. No surprises here, just as in the classical Bessel and Airy cases in \cite{DG}.

\bigskip

\bigskip

In section $3$
we look in detail at a few examples of exceptional Hermite polynomials using the very explicit expressions given in \cite{GUKKM} and obtain some {\bf NEW} ad-conditions which are then compared with the ones one would get by using M. Reach's method. It turns out that the new ad-conditons involve lower powers of the ad-operator acting on $\Theta$. More explicitly: in the case of a recursion relation with $2(k+1)+1$ terms, the relations of M. Reach start with $$A_{2(k+1)+1}$$ while the new ad-conditions that we derive for these examples start with  $$A_{k+2}$$.

\bigskip

In section $4$ we carry out a few Darboux steps starting from the classical Laguerre situation and look for lower order ad-conditions in the case of exceptional Laguerre polynomials.
The results obtained so far are very different from the Hermite case: namely, we
have failed to find any ad-conditions simpler that the ones that follow from M. Reach's general method as extended in this paper.

\bigskip

\bigskip

Section $5$ contains some of the main new results in this paper.
We extend the method given by M. Reach and used by him in a few examples and 
we describe a general method for solving his and similar ad-conditions.
The fact that by solving a few of these equations we find situations not connected with known examples may open the door to new interesting examples enjoying the bispectral property.

\bigskip

\noindent
We obtain the
general solution for the second ad-condition (5) obtained by M. Reach, as well
as the general solution for the simple equations

\begin{equation}
A_2- 4 A_0=0.
\end{equation}

and 

\begin{equation}
A_3-16 A_1=0.
\end{equation}

We also give the general solution to a different equation, namely

\begin{equation}\label{new}
A_4-40 A_2 +144 A_0=0.
\end{equation}

\noindent
since these last three conditions are relevant in the Hermite case. This condition has been reported in \cite{CG24}.

\bigskip

A very important point in obtaining these (and other) general solutions is the fact that, exactly as in (1.29) in \cite{DG}, the equations in the discrete-continuous case gives for $V$ the expression

%$$V= \frac{\partial}{\partial x} \left( P \left( \frac{\partial \Theta}{\partial x} \right)^{\!-1} \right)$$

\begin{equation}
V=\left( \frac{P}{\Theta'} \right)'
\end{equation}

In the same fashion as in \cite{DG}, $P$ and $\Theta$ are polynomials of appropriate degrees.

\bigskip

In section $6$ we write explicitly the recursions that can be obtained with the method indicated by M. Reach in \cite{Reach123}.

\bigskip

In section $7$
we do something similar that applies (apparently) only in the Hermite case.

\bigskip

In section $8$ we display some new ad-conditions that hold when dealing with matrix valued orthogonal polynomials.

\bigskip

In section $9$ we display a consequence of the existence of ad-conditions.

\bigskip

In section $10$ we compare the results in \cite{DG,GH1} with the ones in this paper as well as all the results concerning exceptional orthogonal polynomials.

\bigskip

In section $11$ we make some final comments.

\bigskip

\bigskip

\section{A few examples of exceptional Hermite polynomials}

Using the very nice tools in \cite{GUKKM} we get, for each value of $k=0,1,2,3,..$,
a potential $V$ and a $\Theta$ function as given below.

\bigskip

The differential operator is
$$
\mathcal{L}=-\partial^2_{x}+x^2-2\dfrac{d^2}{dx^2}\log H_{k}(x)
$$

and the eigenvalue is given by  $$\Theta_k(x)=H_{k+1}(x)$$

\noindent
for an appropriate recursion relation of length $2 (k+1)+1$ satisfied by a family of eigenfunctions $\psi_n$ of $\mathcal{L}$. Here $H_k(x)$ denote the standard Hermite polynomials.

\bigskip

A few examples of the {\bf new} ad-conditions are given below. They are independent, and of "lower order" than those that can be obtained by using the ideas of M. Reach.

\bigskip

For $k=0$ and the (traditional) recursion of length $3$ we have the new ad-condition$$A_2- 4 A_0=0$$
while the method of M. Reach gives $$A_3- 4 A_1=0$$.
Notice that this is a consequence of the one above.

\bigskip

For $k=1$ and a recursion of length $5$ we have the new ad-condition

\begin{equation}\label{newad}
A_3-16 A_1=0.
\end{equation}

\noindent
and one could believe that in this case one has the stronger relation $$A_2-16 A_0=0$$ but this is actually false.

\bigskip

Here M. Reach's method gives

\begin{equation}\label{MReach}
A_5-20 A_3+64 A_1=0.
\end{equation}

which follows from the lower order new ad-condition given above.

\bigskip

For $k=2$  and a recursion of length $7$ we have the new ad-condition $$A_4-40 A_2 +144 A_0=0$$

while the method of M. Reach would give

$$A_7-56 A_5 +784 A_3 -2304 A_1=0$$

which, once again, follows from the new ad-conditions.

\bigskip

For $k=3$ and a recursion of length $9$ we have the new ad-condition  $$A_5-80 A_3 +1024 A_1=0$$

while the stronger condition $$A_4-80 A_2+1024 A_0=0$$ does not hold.

\bigskip

The method of M. Reach would give $$A_9-120 A_7+4368 A_5-52480 A_3+147456 A_1=0$$

\bigskip
Finally for $k=4$ the new ad-conditions would read

$$A_6-140 A_4+4144 A_2-14400A_0=0$$

For a general expression of these new ad-conditions in these Hermite cases see section $7$.

\bigskip

One final example in the Hermite case: in the very nice paper \cite{GUKKM}, one also finds in section $6.1$ some multi-step examples. In the case of the partititon $\lambda=(2,2)$, with gaps at degrees $4$ and $5$ the authors display an eleven term recursion. In this case we have checked that the same ad-condition as in the example above holds, with $\Theta=4 x^5+ 15 x$, and

$$
\mathcal{L}=-\partial^2_{x}+x^2-2\dfrac{d^2}{dx^2}\log (4 x^4 +3)
$$

The careful reader will have noticed that, as in all the examples displayed in this section, the relation between $\Theta(x)$ and the $\tau(x)$ function of M. Sato that appears under the logarithm in the expression for $V$ is given by

$$\Theta'=\tau$$.

This is always true in the case of simple zeros, and observation that one finds in \cite{DG}. Cases where the relation is more complicated are also given in \cite{DG}. This has also surfaced in the context of exceptional orthogonal polynomials.

\bigskip

\bigskip

\section{The case of Laguerre exceptional polynomials}

The classical Laguerre operator is given by

 $$(-(d/dx)^2+(x^2/16+((4m^2-1)/4)/x^2)$$

We find it useful to make the replacement

 $$m=-(k^2+4)/4$$  

 to get $$-(d/dx)^2+ x^2/16+(k^4+8k^2+12)/(16 x^2)$$

This operator has a two dimensional space of eigenfunctions for every eigenvalue $-n=0,-1,-2,...$
\bigskip

A basis of this space is (generically on $k$ or $m$) given by

$$\phi_n(x)=x^{m+1/2 }L_n^{m}(-x^2/4) e^{x^2/8}$$ and
$$\psi_n(x)=x^{-m+1/2} {}_1F_1(-n-m,1-m,-x^2/4) e^{x^2/8}$$
involving the generalized Laguerre polynomials and the confluent hypergeometric function respectively.

\bigskip

\bigskip

We have observed earlier that a bispectral situation as the one considered here requires that the potential $V$ be a rational function of $x$. This imposes serious restrictions in the choice of an eigenfunction needed to perform a Darboux step.

\bigskip

Under one application of the Darboux process using the eigenfunction
$$\phi_1$$ of the initial operator, we get one of the form $$-(d/dx)^2 +V$$
where $V$ is given by

\begin{equation}\label{onestep}
{\frac{2}{\left(x+k\right)^2}}+{\frac{2}{\left(x-k\right)^2}}+{\frac{x^2
 -2\,k^2+8}{16}}+{\frac{k^4-4}{16\,x^2}}%\leqno{\tt}
\end{equation}

Notice that if we take away the polynomial part of $V$ to get

$${{2}\over{\left(x+k\right)^2}}+{{2}\over{\left(x-k\right)^2}}+{{k^4
 -4}\over{16\,x^2}}\leqno{\tt}$$

this can be written as

$$-2\,\log \left(x^{{{k^4-4}\over{32}}}\,\left(x^2-k^2\right)\right)\leqno{\tt}$$

\bigskip
The ad-condition in this case reads as follows

$$A_5 - 5 A_3 + 4 A_1=0$$

Another application of the Darboux method, with a carefully chosen eigenfunction of this last operator gives for $V$ the expression

$${{4}\over{x^2-k^2+2\,k}}+{{8\,k^2-16\,k}\over{\left(x^2-k^2+2\,k
 \right)^2}}+{{4}\over{x^2-k^2-2\,k}}+{{8\,k^2+16\,k}\over{\left(x^2-
 k^2-2\,k\right)^2}}+{{x^2-2\,k^2+16}\over{16}}+{{k^4-8\,k^2+12
 }\over{16\,x^2}}\leqno{\tt}$$

\noindent
if we remove from $V$ the polynomial part we obtain a piece that 
can be written as

$$-2\,\log \left(x^{{{k^4-8\,k^2+12}\over{32}}}\,\left(x^2-k^2-2\,k
 \right)\,\left(x^2-k^2+2\,k\right)\right)\leqno{\tt}$$

\bigskip

The corresponding ad-condition is

$$A_7 -14 A_5 +49 A_3 -34 A_1=0$$

One more careful application of the Darboux method gives a new $V$ whose polynomial part is $$(x^2-2 k^2 +24)/16$$ plus a piece that can be written as

%$${{12\,x^4-12\,k^4+336\,k^2}\over{x^6-3\,k^2\,x^4+\left(3\,k^4-12\,k
% ^2\right)\,x^2-k^6+12\,k^4-32\,k^2}}+\\{{\left(288\,k^4+2304\,k^2
% \right)\,x^4+\left(2304\,k^4-576\,k^6\right)\,x^2+288\,k^8-3456\,k^6
% +9216\,k^4}\over{\left(x^6-3\,k^2\,x^4+\\\left(3\,k^4-12\,k^2\right)\,
% x^2-k^6+12\,k^4-32\,k^2\right)^2}}+{{k^4-16\,k^2+60}\over{16\,x^2}}\leqno{\tt}$$

$$-2\,\log \left(x^{{{k^4-16\,k^2+60}\over{32}}}\,\left(x^6-3\,k^2\,x
 ^4+\left(3\,k^4-12\,k^2\right)\,x^2+12\,k^4-32\,k^2-k\right)\right)\leqno{\tt}$$
\bigskip

The ad-condition in this case is

$$A_9 -30 A_7 +273 A_5 -820 A_3 +576 A_1=0$$

The general form of all these ad-conditions 
 obtained using M. Reach's method can be seen in section $6$.

\bigskip
In each one of the examples above one has a bispectral situation, i.e. not only
a second order differential operator but also a recursion relation satisfied by a family of its eigenfunctions. The length of these recursion relations grows as we perform Darboux steps.

\bigskip

Notice that the $\tau$ functions that appear here are similar not some of the ones with multiple zeros that appear already in \cite{DG}.

\bigskip

\section{Solving some ad-conditions}

As mentioned earlier this section contains most of the main results of this paper.
We first describe a method that will be used in the examples considered here.
This part of the argument is exactly the one given in \cite{DG}.

\bigskip

If we are trying to solve for $\Theta$ and $V$ {\bf any} ad-condition starting with $A_n$ (and involving a few more $A_j$ with values of $j$ lower than $n$) we first notice that the coefficient going with the derivative of order $n$ of the argument $f$ in this differential operator, gives that $\Theta$ is a polynomial of degree at most $n-1$. The next coefficient going down with our differential operator gives no additional information.

\bigskip

The next coefficient down contains a lot of new information.
It gives that the derivative of order $n$ of the product $\Theta' \int V$ plus a constant $d_n$ (that depends on $n$) multiplied by the derivative of order $n-2$ of $\Theta$ vanishes. Integrating this $n$ times we conclude that

\begin{equation}
V=\left( \frac{P}{\Theta'} \right)'
\end{equation}

\noindent
where $\Theta$ is a polynomial of degree
at most $n-1$ and $P$ is a polynomial of degree at most $n+1$ with one more
restriction that will appear below.

\bigskip

We now apply this to a few examples, starting with rather simple ones.

\bigskip

First we dispose of two very simple cases, that could be (almost) handled by hand. For the 
remaining cases the facts mentioned above, and the help of a symbol manipulator such as Maxima are highly recommended. 

\bigskip

For the equation $$A_2-4 A_0=0$$
it is not hard to see that the general solution is

$$ \Theta=x , V=x^2+c_{1}$$

which contains as a special case the original Hermite case

$$ \Theta=x , V=x^2 $$

For the equation $$A_3-16 A_1=0$$ once again it is not too hard to see that the general solution is

$$ \Theta=a_{2}\,x^2+a_{1}\,x , V={{4\,a_{2}^2\,x^2+4\,a_{1}
 \,a_{2}\,x+2\,a_{2}\,c_{2}-a_{1}^2}\over{4\,a_{2}^2}}-{{2\,a_{1}^2\,
 a_{2}\,c_{2}+8\,c_{0}\,a_{2}^3-4\,a_{1}\,c_{1}\,a_{2}^2-a_{1}^4
 }\over{4\,a_{2}^2\,\left(2\,a_{2}\,x+a_{1}\right)^2}} \leqno{\tt}$$

which has as a special case, the Hermite example after one application of Darboux's method, namely

$$ \Theta=a_{2}\,x^2 , V=x^2+{{2}\over{x^2}}  \leqno{\tt}$$

We now move to more complicated cases.

\bigskip

We start by looking at the general solution of $$A_5-5 A_3 +4 A_1=0$$.

The operator in question is a differential operator of order
$5$. The top order coefficient gives that $\Theta$ is a polynomial of degree at most $4$. The coefficient of order $4$ is then automatically zero. The next one down
gives the condition (alluded to above)

$$(\Theta'  \int V)^{\romannumeral 5}- (5/2) \Theta'''=0$$ 

\bigskip

\noindent
and then very much as in \cite {DG}, this allows one to write

\begin{equation}
V=\left( \frac{P}{\Theta'} \right)'
\end{equation}

\noindent
where $\Theta$ is, as noticed earlier, a polynomial of degree
at most $4$ and $P$ is a polynomial of degree at most $6$ with one more
restriction that will appear below.

\bigskip

From the observations above we can put

\begin{equation}
\Theta=\sum_{i=1}^{4} a_{i} x^i
\end{equation}

and

\begin{equation}
P=\sum_{i=0}^{6} c_{i} x^i
\end{equation}

\noindent
and from the third coefficient, down from the top one of the differential operator, we obtain as a consequence of the
extra restriction alluded to above the fact that $$c_6=a_4/12, c_5=a_3/8$$

	Notice that we can assume that $\Theta(0)=0$; adding a constant to $\Theta(x)$ does not affect the ad-conditions.

\bigskip

Going on to analyze carefully the remaining coefficients of our differential operator we get that
the list of possible solutions is given below
\bigskip

A first solution given by

$$\Theta=x, V=x^2+c x$$

A second solution given by

$$\Theta=x, V=x^2/4 +c x$$

A third solution given by

$$\Theta=x^2+c x, V=x^2/4 + e  x$$

Here and below, quantities such as $c,e,c_1,p_1,a,b$ are arbitrary constants.

\bigskip

A fourth solution given by

$$\Theta=x^2+c_{1}\,x$$

$$V={{4\,p_{1}}\over{\left(2\,x+c_{1}\right)^2}}+{{x^2+c_{1}\,x
 }\over{16}}$$

A fifth solution given by

$$\Theta={{x\,\left(2\,x+a\right)\,\left(4\,x^2+2\,a\,x+4\,b-a^2
 \right)}\over{8}}\leqno{\tt}$$

$$V={{2\,x^2+a\,x}\over{32}}\leqno{\tt}$$

A sixth solution given by

$$\Theta=x\,\left(x+2\,e_{1}\right)\,\left(x^2+2\,e_{1}\,x-4\,
 e_{1}^2+b\right)\leqno{\tt}$$

$$V={{p_{1}}\over{\left(x+e_{1}\right)^2}}+{{x^2+2\,e_{1}\,x}\over{16
 }}\leqno{\tt}$$

\noindent
and the most interesting solution for the potential $V$ is given by the sum of  four terms 
expressed in terms of the coefficients of $\Theta$, namely

$${{2}\over{\left(x+{{\sqrt{3\,a_{3}^2-8\,a_{2}\,a_{4}}+a_{3}}\over{4
 \,a_{4}}}\right)^2}} +$$

$${{2}\over{\left(x-{{\sqrt{3\,a_{3}^2-8\,a_{2}\,a_{4}}-a_{3}}\over{4
 \,a_{4}}}\right)^2}} +$$

$$-{{1024\,a_{4}^4-64\,a_{2}^2\,a_{4}^2+48\,a_{2}\,a_{3}^2\,a_{4}-9\,
 a_{3}^4}\over{4096\,a_{4}^4\,\left(x+{{a_{3}}\over{4\,a_{4}}}\right)
 ^2}} +$$

$${{48\,a_{4}^2\,x^2+24\,a_{3}\,a_{4}\,x+192\,a_{4}\,c_{4}-8\,a_{2}\,
 a_{4}-9\,a_{3}^2}\over{768\,a_{4}^2}}$$

\bigskip

By means of a shift in the variable $x$ we could have assumed $a_3=0$.
Notice that if (having set $a_3=0$) we put $k=\sqrt{-a_2/(2 a_4)}$
then the sum above differ in an unimportant additive constant from the potential obtained after one step of Darboux and given in section $4$.

\bigskip

\bigskip

\bigskip

Now we consider the equation

$$A_4-40 A_2 +144 A_0=0$$ 

\noindent
which appeared in connection with a Hermite type example given earlier.

\bigskip

Proceeding as before we get that $\Theta$ is a polynomial of degree at most $3$ and $P$ is a polynomial of degree at most $5$. The third coefficient starting from the top gives now the relation

$$(\Theta'  \int V)^{\romannumeral 4}-20 \Theta''=0$$

We can thus assume

\begin{equation}
\Theta=\sum_{i=1}^{3} a_{i} x^i
\end{equation}

and

\begin{equation}
P=\sum_{i=0}^{5} c_{i} x^i
\end{equation}

\noindent
and from this third coefficient from the top we obtain the
extra restriction alluded to above, namely $$c_5=a_3, c_4=(5/3) a_2$$

By continuing carefully down all the coefficients of our differential operator , the list of all solutions is given by

\bigskip
A first solution given by $$\Theta=x, V=9 x^2+c$$
A second solution given by $$\Theta=x,V=x^2+c$$
A third solution given by $$\Theta=x^3,V=x^2+c$$
A fourth solution given by $$\Theta=x^3,V=x^2+c+2/x^2$$
A fifth solution given by

$$ \Theta=x^3+a_{1}\,x , V={{9\,x^2+3\,c_{3}-a_{1}}\over{9}}
   \leqno{\tt }$$

A sixth solution given by $$\Theta=x^3+(3/2) x,V=2/(x-i/\sqrt 2)^2+2/(x+i/\sqrt 2)^2+x^2+c$$
A seventh solution given by  $$\Theta=x^3-(3/2) x,V=2/(x-1/\sqrt(2))^2+2/(x+1/\sqrt(2))^2+x^2+c$$

This solution corresponds to the Hermite type example given earlier after two steps of the Darboux process.

\bigskip

An eighth solution given by $$\Theta=a_3 x^3+a_2 x^2+ (2/9) a_2^{2}/a_3,V=x^2+(2/3) a_2/a_3 x$$ 
A ninth solution given by $$\Theta=x^3+3^{3/2} x^2+  2^{1/2} 3 x$$
	and V consisting of the sum of three parts, namely

$$x^2+\sqrt{2}\,\sqrt{3}\,x+{{c_{3}-10\,a_{3}}\over{3\,a_{3}}}$$

$${{2}\over{\left(x+{{\sqrt{3}+3}\over{\sqrt{2}\,\sqrt{3}}}\right)^2
 }}$$

\noindent
and

$${{2}\over{\left(x-{{\sqrt{3}-3}\over{\sqrt{2}\,\sqrt{3}}}\right)^2
 }}$$

By shifting $x$ by $\sqrt{3/2}$ this solution becomes the seventh solution above.

\bigskip

Finally there is an extra solution given by

	$$\Theta=x^3-3^{3/2} x^2+ 2^{1/2} 3 x$$
and V consisting of the sum of two parts, namely a polynomial part
$$x^2-\sqrt{2}\,\sqrt{3}\,x+{{c_{3}-10\,a_{3}}\over{3\,a_{3}}}\leqno{\tt}$$

plus the {\bf ratio} of

$$2^{{{5}\over{2}}}\,3^{{{23}\over{2}}}\,a_{3}\,\left(x+{{\sqrt{3}\,i
 -3}\over{\sqrt{2}\,\sqrt{3}}}\right)\,\left(x-{{\sqrt{3}\,i+3}\over{
 \sqrt{2}\,\sqrt{3}}}\right)$$

and

$$\sqrt{2}\,3^{{{23}\over{2}}}\,a_{3}\,\left(x-{{\sqrt{2}\,\sqrt{3}-
 \sqrt{2}}\over{2}}\right)^2\,\left(x-{{\sqrt{2}\,\sqrt{3}+\sqrt{2}
 }\over{2}}\right)^2$$

This last expression for $V(x)$ is very different from other solutions listed above and deserves further study.
It is also clear that some of these solutions can be simplified by a shift of variables.
\bigskip

\bigskip

\bigskip

\section{An explicit form of Reach's ad-conditions} 

A compact from for M. Reach's conditions when dealing with a recursion of length $2 n+1$ is given by

%$$(({ad}\mathcal{L}^{2} -n^2)
% ({ad}\mathcal{L}^{2} -(n-1)^2)...
% ({ad}\mathcal{L}^{2} -4)
% ({ad}\mathcal{L}^{2} -1) ({ad}\mathcal{L}))(\Theta)=0.$$

\bigskip

\[
	\prod_{i=1}^{n}({ad}\mathcal{L}^{2}-i^{2}) {ad}\mathcal{L} (\Theta)=0;
\]

\section{An explicit form for the new ad-conditions for some Hermite exceptional polynomials}

We find that there are two forms for these conditions depending on the parity of $k$. Recall that we are dealing with recursion relations of length $2(k+1)+1$.

\bigskip

When $k$ is odd, and thus $\Theta(x)=H_{k+1}(x)$ we have

%$$(({ad}\mathcal{L}^{2} -(4(k-2))^2)
% ({ad}\mathcal{L}^{2} -(4(k-3))^2)...
% ({ad}\mathcal{L}^{2} -(4 (2))^2)
% ({ad}\mathcal{L}^{2} -(4 (1))^2)) ({ad}\mathcal{L}))(\Theta)=0.$$

\bigskip

\[
	\prod_{i=1}^{(k+1)/2}({ad}\mathcal{L}^{2}-(4 i)^{2}) {ad}\mathcal{L} (\Theta)=0;
\]
\bigskip

When $k$ is even , and thus once again $\Theta(x)=H_{k+1}(x)$ we have

%$$(({ad}\mathcal{L}^{2} -((2+(4(k-3)))^2))
% ({ad}\mathcal{L}^{2} -((2+(4(k-4)))^2))...
% ({ad}\mathcal{L}^{2} -((2+4 (1))^2))
% ({ad}\mathcal{L}^{2} -((2+4 (0))^2)) )(\Theta)=0.$$

\[
	\prod_{i=0}^{k/2}({ad}\mathcal{L}^{2}-(2+4 i)^{2}) (\Theta)=0;
\]
\bigskip

\bigskip

\section{A peek at the non-commutative case}

A basic reference for this section is \cite{CG8}, as well as a contribution by A. Duran included in the book \cite{MOURAD}. A. Duran has a nice catalog of examples, and the same book contains a contribution by I. Pacharoni, J. A. Tirao and myself.

We have looked at a few examples dealing with matrix valued orthogonal polynomials (of the Hermite and Laguerre type) whose weights are given in the references just mentioned.

\bigskip

In the case of Hermite type matrix valued weights we have found a new form of the ad-conditions involving {\bf matrix valued universal coefficients} of the kind that appeared in expression (3).

\bigskip
We report here on one example, but we have checked this in several other ones

\begin{equation}
\mathcal{L} F=	F''(t)+F'(t)
\begin{pmatrix}
-2t & 2a\\
0 & -2t
\end{pmatrix}
+F
\begin{pmatrix}
-2 & 0\\
0 & 0
\end{pmatrix}
\end{equation}

\bigskip

In this case the ad-conditions become

$$A_2 - 4 A_0=0$$

with $$\Theta=x I$$.

\bigskip

\bigskip

In the case of Laguerre type matrix valued weights, we display a couple of examples of these new kind of ad-conditions, although we have checked this in a few more cases

\begin{equation}
\mathcal{L} F=	F''(t)+F'(t)
\begin{pmatrix}
-2t & 4at\\
0 & -2t
\end{pmatrix}
+F
\begin{pmatrix}
-4 & 2a\\
0 & 0
\end{pmatrix}
\end{equation}

\noindent
and in this case we have 

\begin{equation}
(A_2-4 A_0)=0 
=0
\end{equation}

\noindent
with $\Theta=x I$.

\bigskip

\noindent
A second Laguerre example is given by

\begin{equation}
\mathcal{L} F=	F''(t)+F'(t)
\begin{pmatrix}
2b-2t & 2a-2abt\\
0 & -2t
\end{pmatrix}
+F
\begin{pmatrix}
-2 & 0\\
0 & 0
\end{pmatrix}
\end{equation}

In this case the ad-conditions become

$$A_2 M_2 - 4 A_0 M_0=0$$

with $$\Theta=x I,\quad M_2=\begin{pmatrix}
	0 & 0\\
	r_2  & r_1
\end{pmatrix}, \quad M_0=\begin{pmatrix}
	a b r_2 & a b r_1\\
	r_2  & r_1
\end{pmatrix}.$$

\bigskip
Notice that we have a set of different ad-conditions, by choosing the arbitrary parameters $r_1,r_2$.

\bigskip

\bigskip

\section{The ad-conditions and the Heisenberg evolution of $\Theta(x)$}

Using the well known Baker-Campbell-Hausdorff expression from Lie algebras, familiar to physicists as the "Heisenberg interaction picture", 

$$ e^{t \mathcal{L}} \Theta(x) e^{-t \mathcal{L}} =
\sum_{i=0}^{\infty} \frac{t^i}{i!}  A_i$$

\bigskip

we explore a consequence of relations like (7) or (9) above.

\bigskip

In the case of $k=1$ of section $3$ and using $A_3=16 A_1$ and its obvious consequences one easily gets that the expression above in this exceptional Hermite case becomes

$$ e^{t \mathcal{L}} \Theta(x) e^{-t \mathcal{L}} =
\cosh(4 t) A_0+ \sinh(4 t) A_1/4$$ 

In this simple case this expression could be checked by using Mehler's expression for the fundamental solution of the Schr\"odinger operator going with the harmonic oscillator. Such a fundamental solution is, to the best of my knowledge, not known when the harmonic oscillator is subject to a Darboux transformation.
%The value of $A_0$ is $\Theta$ by definition and $A_1=\Theta  \Theta' +2 (\Theta')

\section{Acquiring a degree of freedom after a step of the Darboux process}

Anyone comparing the situation in \cite{DG,GH1,GRUHAI,Haine1,Haine77} with the situation here could be left wondering about the lack of free "time" parameters. In \cite{DG} we can use the entire two dimensional space of eigenfunctions of our second order differential operators when performing the Darboux process. In \cite{GH1,GRUHAI} we use the full two dimensional space of eigenfunctions of the bi-infinite tridiagonal matrices involved. In \cite{GRUNHAIN} we consider the full two dimensional space of eigenfunctions and in this way generalize the Askey-Wilson polynomials.
Both in the Laguerre, as well as in the Hermite case, see \cite{CG7}, one has to be careful in picking eigenfunctions so that the Darboux process will produce rational potentials, and we do not have the degrees of freedom that appear as "times" for certain nonlinear evolutions in the earlier cases.

\section{Some remarks on the literature and some final thoughts}

Just as in \cite{DG}. M. Reach found some instances of the bispectral propery by solving his ad-conditions directly. A later paper by M. Reach, \cite{R} shows the relevance of the Darboux process in his continuous-discrete case. In Reach's work there is no restriction to polynomial eigenfunctions, and at a later time Christiane Quesne, see \cite{Qu}, appears to be the first one to point out that the Darboux process is very useful when dealing with "exceptional polynomials". Two remarks about the very nice paper by C. Quesne: she cites the groundbreaking paper of
 V. Bargmann, \cite{Barg} which got the ball rolling, and in a very generous way cites the use of the Darboux process in \cite{DG}. In this regard I would like to draw attention to the little-noticed paper \cite{R}, where the Darboux process is heavily used. In this very nice paper, M. Reach makes heavy use  of Wronskians, using tools given in \cite{AdMos} as well as extending some old results of Jacobi. 

\bigskip

A pre-historical remark: the Darboux process is
 due to Moutard, see \cite{Dar}. This is given in Livre IV, Chap IX, No. 408, and the attribution to Moutard is given in Chap II, No. 343, page 53 of \cite{Dar}. This powerful method may be even older, see \cite{Gesz}, and it has been rediscovered several times, including by E. Schr\"odinger himself, see \cite{Schr}.

\bigskip

We have already pointed out that \cite{DG} showed a surprising connection between the Korteweg-deVries flow and the bispectral problem. This was further illustrated in \cite{ZM} as mentioned earlier in this paper.
Other nonlinear flows, related to the continuous-discrete version of the bispectral problem, such as the Toda flow have made their appearance too.
Notice that here one needs to use the full two-dimensional eigenspace of our tridiagonal matrix giving the recursion relation.
The first instance of this may be the paper \cite{GRUHAI} which includes the Toda and the Virasoro flows, as well as \cite{Haine1}, see also the very nice paper \cite{Haine77}.

\bigskip

The original motivation to pose the bispectral problem in \cite{DG} was the issue of ``time-and-band-limiting'' as motivated by medical imaging in the case of limited and noisy data. This is mentioned in the introduction to \cite{DG}. This has been extended beyond the Fourier case, involving the second derivative, to operators obtained from it by a few Darboux transformations, see for instance \cite{CGYZ4} and some of its references.

The prolate differential operator that commutes with the integral one of "time-and-band limiting"  has surfaced in rather unexpected places: for a remarkable instance of this see \cite{CM}.
If one wants to get a fuller picture of the large collection of papers linked to these mathematical developments, one could look at the small sample of papers \cite{BHY1,BHY2,Boch,BurCha,CGYZ1,CGYZ3,CG7,Crum,Deift,GGU2,G888,G6,Haine1,HK,KR,KM,STZ,W2}
and some of the references in these papers.

\bigskip

One could say that its relevance in the "time-and-band limiting" problem, and the search of differential operators commuting with integral ones is a good application of the bispectral poblem, but this is in my view a bit of a stretch. I would rather argue that the reach of the "time-and-band limiting" problem is so broad (as for instance in \cite{CM}) that it has also touched another nice piece of mathematics, namely the bispectral problem.

\bigskip

If one keeps in mind that we are looking for situations that extend the time honored cases of Hermite, Laguerre, Jacobi, Bessel and Airy who grew out and have been an important tool in mathematical physics for a very long time, I think that it is fair to say that these more recent tools, have not yet found deep applications.
Looking for methods such as solving explicitly the ad-conditions both in the scalar and the non-commutative cases could help in this important task.  

%\end{abstract}

%\maketitle

\section{Statements and Declarations}

The corresponding (and only) author states that there is no conflict of interest and data sharing is not applicable to this article as no datasets were generated or analyzed during the current study. 

%\end{Statements and Declarations}

\end{document}